\documentclass{aip-cp}

\usepackage[numbers]{natbib}
\usepackage{rotating}
\usepackage{graphicx}
\usepackage{xcolor}

\begin{document}

\title{Recent Results of the \textsc{Majorana Demonstrator} Experiment}

\author[aff1]{J.M.L\'opez-Casta\~no\corref{cor1}}
\author[aff2]{S.I.~Alvis}	
\author[aff3]{I.J.~Arnquist} 
\author[aff4,aff5]{F.T.~Avignone~III}
\author[aff6]{A.S.~Barabash}
\author[aff1]{C.J.~Barton}	
\author[aff7]{V.~Basu} 
\author[aff5]{F.E.~Bertrand}
\author[aff8]{B.~Bos} 
\author[aff9]{V.~Brudanin}
\author[aff10,aff11]{M.~Busch}	
\author[aff2]{M.~Buuck}
\author[aff12,aff11]{T.S.~Caldwell}
\author[aff13]{Y-D.~Chan}
\author[aff8]{C.D.~Christofferson} 
\author[aff14]{P.-H.~Chu} 
\author[aff12,aff11]{M.L.~Clark}
\author[aff2]{C.~Cuesta\footnote{Present address: Centro de Investigaciones Energ\'{e}ticas, Medioambientales y Tecnol\'{o}gicas, CIEMAT 28040, Madrid, Spain}\textcolor{white}{.}$^{,}$}
\author[aff2]{J.A.~Detwiler}
\author[aff13]{A.~Drobizhev}
\author[aff4]{D.W.~Edwins}
\author[aff15,aff5]{Yu.~Efremenko}
\author[aff16]{H.~Ejiri}
\author[aff14]{S.R.~Elliott}
\author[aff12,aff11]{T.~Gilliss}
\author[aff17]{G.K.~Giovanetti} 
\author[aff18,aff11,aff5]{M.P.~Green}
\author[aff19]{J.~Gruszko} 
\author[aff2]{I.S.~Guinn}		
\author[aff4]{V.E.~Guiseppe}	
\author[aff12,aff11]{C.R.~Haufe}
\author[aff12,aff11]{R.J.~Hegedus}
\author[aff12,aff11]{R.~Henning}
\author[aff12,aff11]{D.~Hervas~Aguilar} 
\author[aff3]{E.W.~Hoppe}
\author[aff2]{A.~Hostiuc}
\author[aff12,aff11]{M.A.~Howe} 
\author[aff20]{K.J.~Keeter}
\author[aff21]{M.F.~Kidd}
\author[aff6]{S.I.~Konovalov}
\author[aff3]{R.T.~Kouzes}
\author[aff15]{A.M.~Lopez}	
\author[aff12,aff11]{E.~Martin}	
\author[aff7]{R.D.~Martin}	
\author[aff14]{R.~Massarczyk}		
\author[aff12,aff11]{S.J.~Meijer}	
\author[aff22,aff23]{S.~Mertens}
\author[aff13]{J.~Myslik}		
\author[aff1]{T.K.~Oli}  
\author[aff12,aff11]{G.~Othman}
\author[aff2]{W.~Pettus}	
\author[aff7]{A.~Piliounis}
\author[aff13]{A.W.P.~Poon}
\author[aff5]{D.C.~Radford}
\author[aff12,aff11]{J.~Rager}
\author[aff12,aff11]{A.L.~Reine}	
\author[aff14]{K.~Rielage}
\author[aff2]{N.W.~Ruof}
\author[aff5]{B.~Shanks}	
\author[aff9]{M.~Shirchenko}
\author[aff4]{D.~Tedeschi}		
\author[aff5]{R.L.~Varner} 
\author[aff9]{S.~Vasilyev}
\author[aff14]{B.R.~White}
\author[aff12,aff11,aff5]{J.F.~Wilkerson}
\author[aff2]{C.~Wiseman}
\author[aff1]{W.~Xu}
\author[aff9]{E.~Yakushev}
\author[aff5]{C.-H.~Yu}
\author[aff6]{V.~Yumatov}
\author[aff9]{I.~Zhitnikov} 
\author[aff14]{B.X.~Zhu} 

\affil[aff1]{Department of Physics, University of South Dakota, Vermillion, SD 57069, USA}  
\affil[aff2]{Center for Experimental Nuclear Physics and Astrophysics, and Department of Physics, University of Washington, Seattle, WA 98195, USA}
\affil[aff3]{Pacific Northwest National Laboratory, Richland, WA 99354, USA}
\affil[aff4]{Department of Physics and Astronomy, University of South Carolina, Columbia, SC 29208, USA}
\affil[aff5]{Oak Ridge National Laboratory, Oak Ridge, TN 37830, USA}
\affil[aff6]{National Research Center ``Kurchatov Institute'' Institute for Theoretical and Experimental Physics, Moscow, 117218 Russia}
\affil[aff7]{Department of Physics, Engineering Physics and Astronomy, Queen's University, Kingston, ON K7L 3N6, Canada}
\affil[aff8]{South Dakota School of Mines and Technology, Rapid City, SD 57701, USA}
\affil[aff9]{Joint Institute for Nuclear Research, Dubna, 141980 Russia}
\affil[aff10]{Department of Physics, Duke University, Durham, NC 27708, USA}
\affil[aff11]{Triangle Universities Nuclear Laboratory, Durham, NC 27708, USA}
\affil[aff12]{Department of Physics and Astronomy, University of North Carolina, Chapel Hill, NC 27514, USA}
\affil[aff13]{Nuclear Science Division, Lawrence Berkeley National Laboratory, Berkeley, CA 94720, USA}
\affil[aff14]{Los Alamos National Laboratory, Los Alamos, NM 87545, USA}
\affil[aff15]{Department of Physics and Astronomy, University of Tennessee, Knoxville, TN 37916, USA}
\affil[aff16]{Research Center for Nuclear Physics, Osaka University, Ibaraki, Osaka 567-0047, Japan}
\affil[aff17]{Department of Physics, Princeton University, Princeton, NJ 08544, USA}
\affil[aff18]{Department of Physics, North Carolina State University, Raleigh, NC 27695, USA}
\affil[aff19]{Department of Physics, Massachusetts Institute of Technology, Cambridge, MA 02139, USA}
\affil[aff20]{Department of Physics, Black Hills State University, Spearfish, SD 57799, USA}
\affil[aff21]{Tennessee Tech University, Cookeville, TN 38505, USA}
\affil[aff22]{Max-Planck-Institut f\"{u}r Physik, M\"{u}nchen, 80805 Germany}
\affil[aff23]{Physik Department and Excellence Cluster Universe, Technische Universit\"{a}t, M\"{u}nchen, 85748 Germany}
\corresp[cor1]{Mariano.Lopez@usd.edu}

\maketitle

\begin{abstract}
The \textsc{Majorana Demonstrator} is searching for neutrinoless double-beta decay in $^{76}$Ge with two modular arrays of natural and $^{76}$Ge-enriched germanium detectors. It is located at the 4850' level of Sanford Underground Research Facility in Lead, South Dakota, USA, and its total mass of germanium detectors is 44.1 kg, of which 29.7 kg is enriched. The analysis of the first 26 kg-yr of data provides an unprecedented energy resolution of 0.13\% in the region of interest at 2039 keV and a background level of 15.4 $\pm$ 2.0 counts/(FWHM t yr). It establishes the lower limit of the half-life of neutrinoless double beta decay as 2.7$\cdot$10$^{25}$ yr in $^{76}$Ge at 90\% CL. This analysis will be summarized here with an emphasis on the energy determination.

\end{abstract}

\section{INTRODUCTION}

\hspace{0.65cm}The \textsc{Majorana Demonstrator} \cite{MJD} uses P-type Point Contact High Purity Germanium (PPC HPGe) detectors to search for neutrinoless double-beta decay (0$\nu\beta\beta$) in $^{76}$Ge. The \textsc{Majorana Demonstrator} is located at the 4850' level of the Sanford Underground Research Facility (SURF) in Lead, SD, USA. It is composed of 58 PPC HPGe detectors divided into 2 cryostats with 7 strings each. Each string is an assembly of 3, 4 or 5 detectors. The total mass of PPC HPGe crystals is 44.1 kg, 29.7 kg of which is enriched to 88\% $^{76}$Ge.

\section{ENERGY}

\hspace{0.65cm}The electrical signals produced by energy depositions in the detectors are sampled by the GRETINA Digitizer Module at 100MHz \cite{GRETINA}, generating digitized waveforms. An example of a typical waveform is in Fig. \ref{Fig_E} where the main features are indicated: the start time (t$_{0}$), the rising edge, and the falling edge. The waveform is used to determine the raw energy with an uncalibrated energy scale, which is then calibrated using calibration data taken with $^{228}$Th sources \cite{Calibration}.

\subsection{Energy determination}

\begin{figure}[h!]
  \includegraphics[height=5cm]{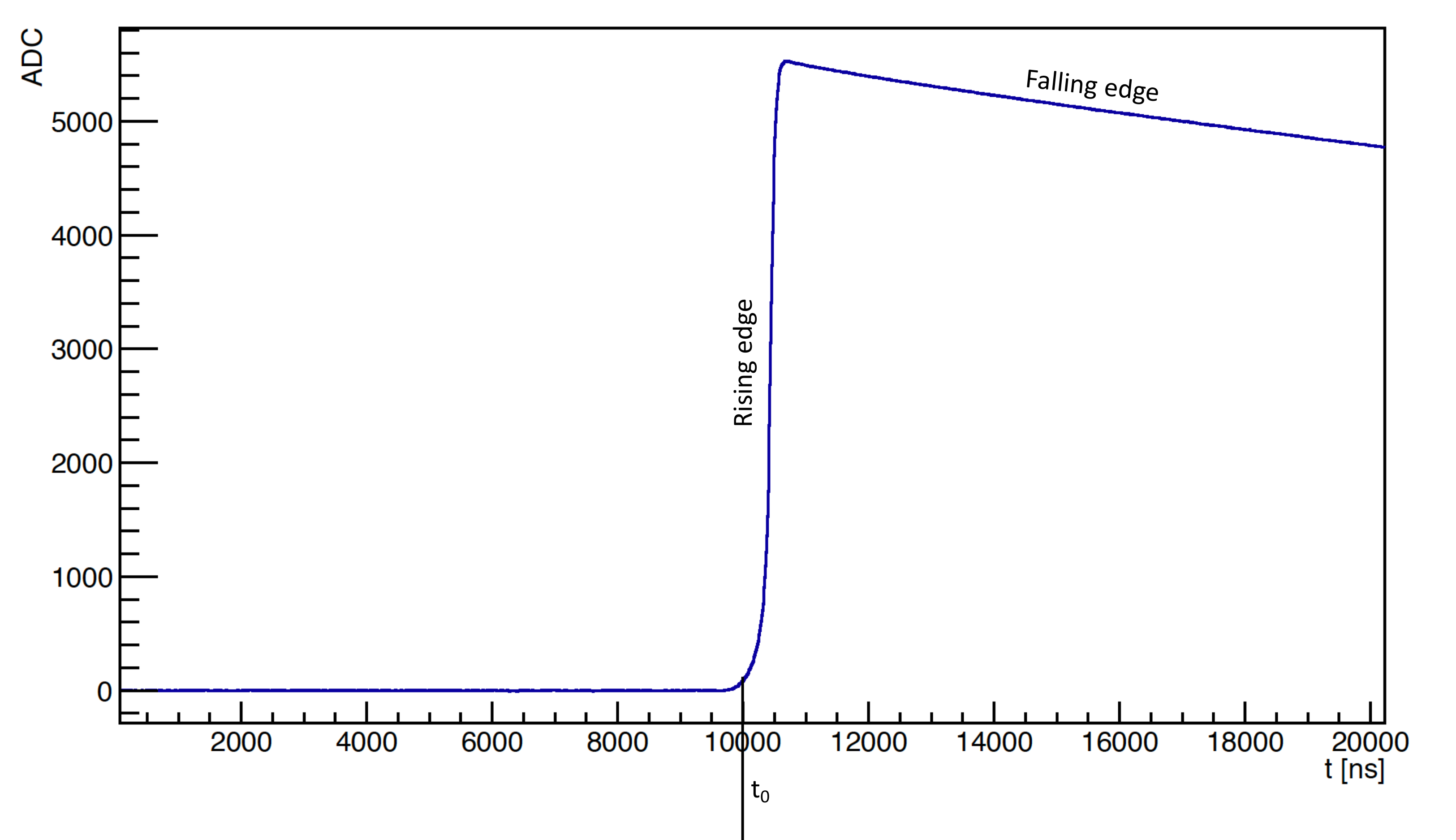}
  \includegraphics[height=5cm]{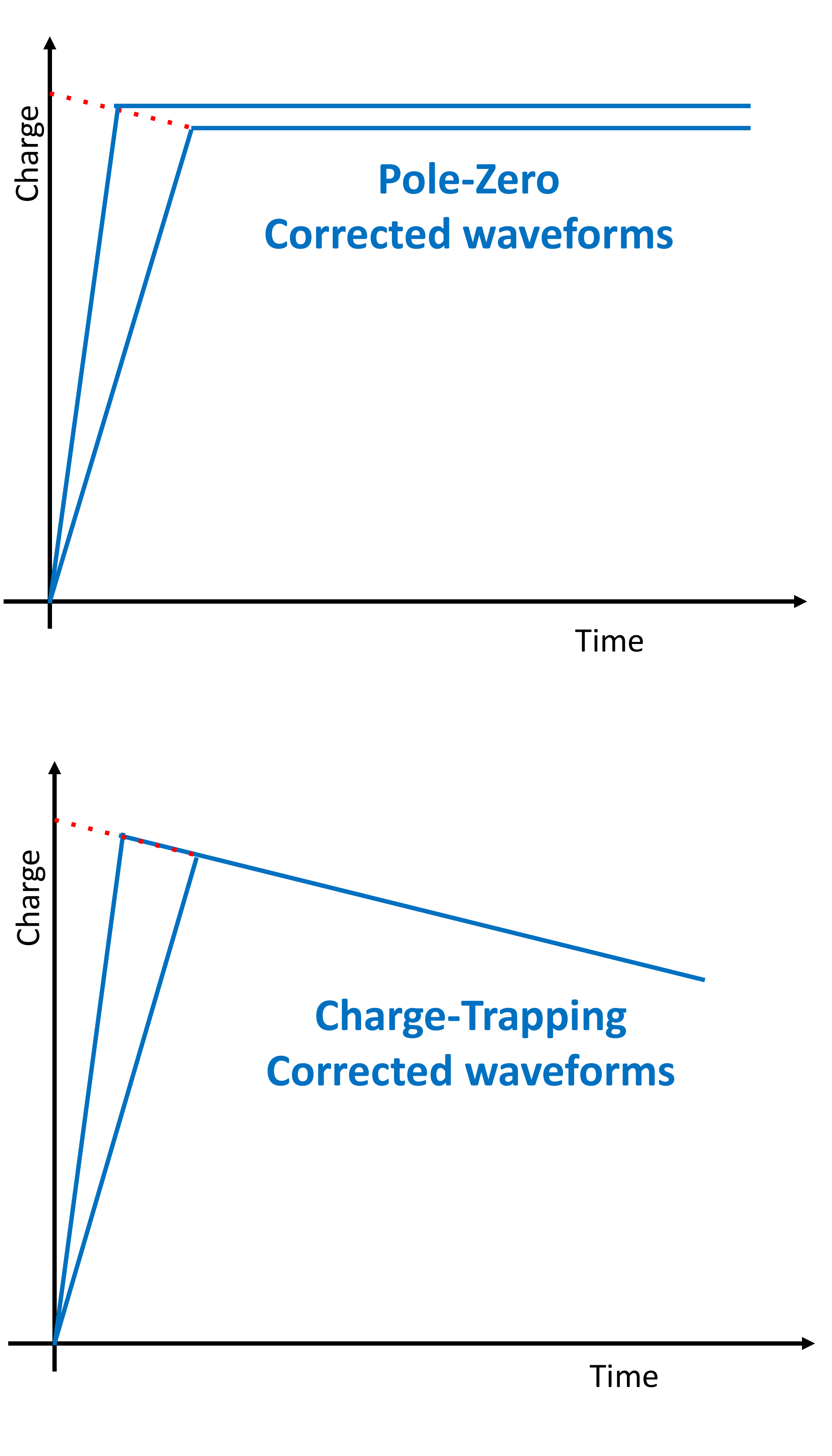}
  \caption{ 
On the left, a digitalized waveform from data, where the start time (t$_{0}$), the rising edge and the falling edge are indicated. On the right, schemes of the expected waveforms for energy depositions in different positions. On the top, the scheme after pole-zero correction, and on the bottom, the scheme after pole-zero and charge trapping corrections.}
  \label{Fig_E}
\end{figure}

\hspace{0.65cm}For each event, the raw energy is estimated using standard Ge detector techniques that measure the amplitude of trapezoidal-filtered \cite{Filtered}, pole-zero corrected  \cite{PZ} signals. The right part of Fig. \ref{Fig_E} includes a scheme of the waveform shape after pole-zero correction. The falling edge of the pole-zero corrected waveform is flat and its value should correspond to the amount of collected charge. However, events of the same energy can have (slightly) different values of collected charge due to the drift-path-dependent charge trapping, providing a degraded energy resolution \cite{CT}. To account for charge trapping effects, the time constant used in the standard pole-zero correction is modified as 1/$\tau$ = 1/$\tau_{PZ}$ + 1/$\tau_{CT}$, where $\tau_{PZ}$ is the standard pole-zero time constant and $\tau_{CT}$ is the time constant associated with charge trapping. While the standard pole-zero correction can be obtained by studying the response of the electronic system, $\tau_{CT}$ is chosen to be the value that minimizes FWHM. After including $\tau_{CT}$, the falling edges are aligned for events with the same energy, as in Fig. \ref{Fig_E}. The raw energy value is evaluated at a fixed time pickoff from t$_{0}$ of the trapezoidal filtered signal after correction.

\subsection{Energy calibration}

\begin{figure}[h!]
  \includegraphics[height=5cm]{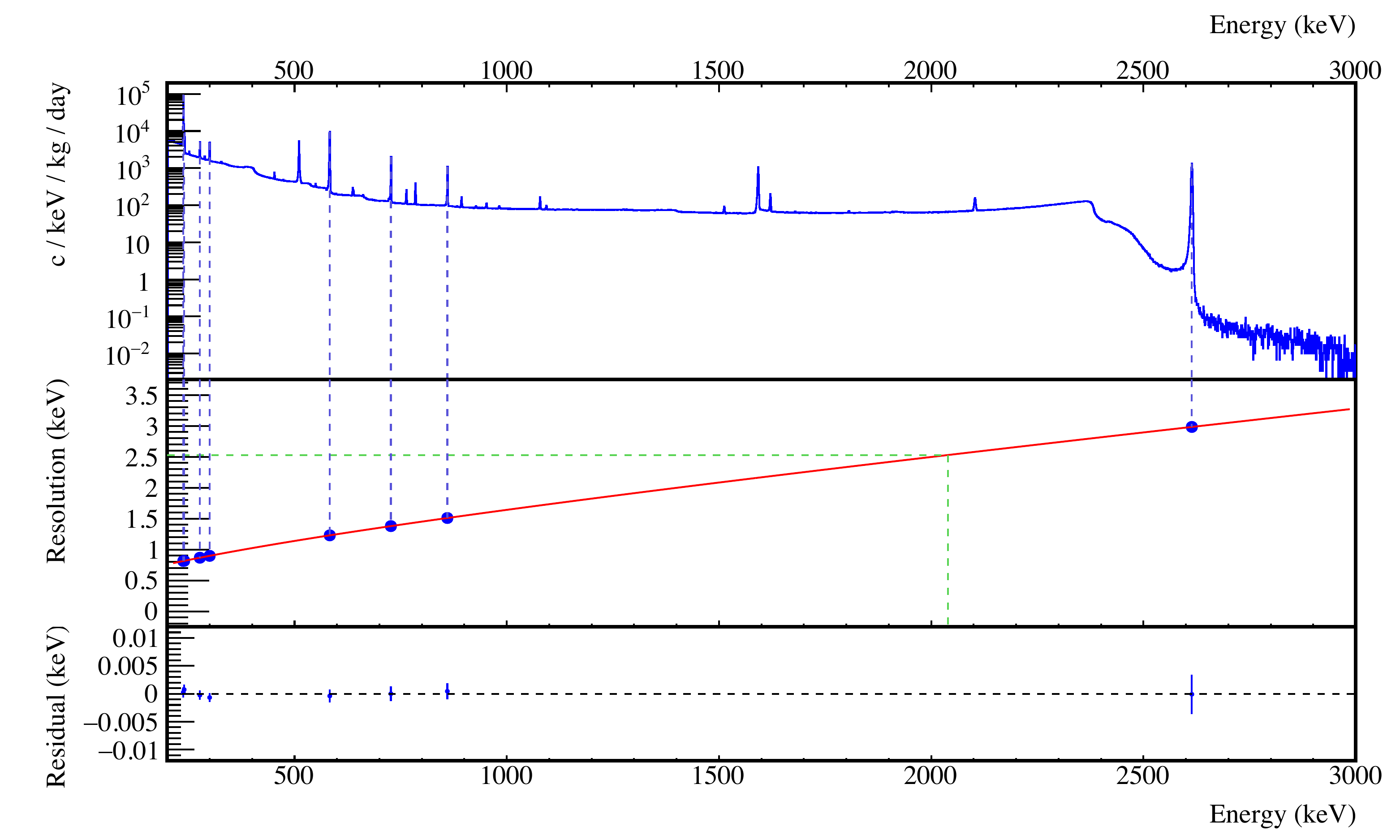}
  \caption{ 
The top panel shows the combined energy spectrum from calibration data. Vertical lines indicates the gamma lines used for the final energy calibration. The center shows the exposure-weighted resolution for each gamma peak used in the calibration and a fit to the exposure-weighted values. The horizontal green line indicate the exposure-weighted average resolution of 2.53 keV at 2039 keV. The bottom shows the residuals from the fit in the center \cite{26MJD}.}
    \label{Fig_B}
\end{figure}

\hspace{0.65cm}This raw energy is converted into keV by energy calibration. Weekly, \textsc{Majorana Demonstrator} takes around 3 hours of data with $^{228}$Th sources inserted into the shield (90 minutes for each module) \cite{Calibration}. This data is used to calibrate each detector individually. The calibration process has two steps. In the first step, an initial linear energy scale is determined from electronic noise at zero energy and the position of the $^{208}$Tl 2614 keV photopeak in the weekly calibration data. In the second step, combined statistics of all calibrations in a subset of data are used to perform a simultaneous linear fit to eight photopeaks in the calibration spectrum. An improved energy calibration is obtained, as shown in Fig. \ref{Fig_B} with the eight peaks indicated. These peaks are at 239 keV ($^{212}$Pb), 241 keV ($^{224}$Ra), 277 keV ($^{208}$Tl), 300 keV ($^{212}$Pb), 583 keV ($^{208}$Tl), 727 keV ($^{212}$Bi), 861 keV ($^{208}$Tl) and 2614 keV ($^{208}$Tl). Peaks due to single-escape and double-escape events, peaks with low amplitude and peaks in close proximity are excluded due to potential differences in peak shape.

\vspace{0.4cm}
Finally, the energy resolution is obtained from combined calibration spectra in each subset. The Full Width at Half Maximum (FWHM) of each peak is determined numerically. Its value is a function of the energy, which fitted the FWHM values  to the square root of a second order polynomial in energy, as Fig. \ref{Fig_B} and the next equation show: $FWHM = \sqrt{}$ ( $\Gamma_{n}^{2}+\Gamma_{F}^{2}E+\Gamma_{q}^{2}E^{2}$), where $\Gamma_{n}$ is the electronic noise term,  $\Gamma_{F}$ is the Fano factor \cite{Fano} term and  $\Gamma_{q}$ is the incomplete charge collection term. The exposure-weighted average of the FWHM at 2039 keV (Q$_{\beta\beta}$($^{76}$Ge)) is 2.53 $\pm$ 0.08 keV, currently the best resolution of all 0$\nu\beta\beta$ experiments at the Q-value.

\section{BACKGROUND REDUCTION}

\hspace{0.65cm}Not all recorded waveforms have been used to search for 0$\nu\beta\beta$. First, they are filtered by a data cleaning process to remove non-physical waveforms while retaining $>$99.9\% of true physics events \cite{DataQ}. As the 0$\nu\beta\beta$ event would not be produced by muons, a muon veto cut is applied to reject events within 1s of a triggering of the muon veto system \cite{Muon}. 0$\nu\beta\beta$ is a single-detector and spatially single-site process since the emitted electrons deposit their total energy within a small range compared with the ability of the detector to localize the energy deposition. This feature is used to reject a range of backgrounds, including multiple Compton-scattered photons, with two different cuts. The first cut is the multiplicity cut, which rejects any events that trigger more than one detector within a $\sim$4 $\mu$s time window. The inefficiencies of this cut and the muon veto are negligible but the dead times are taken into account in the exposure calculation. The other cut, known as AvsE  \cite{MultiSite}, rejects events with multiple energy depositions in the same detector based on the relationship between the maximum current (A) and energy (E). This cut is tuned using $^{228}$Th calibration source data, where 90\% of single-site events are retained. Furthemore, a delayed charge recovery (DCR) cut removes alpha particles that impinge upon the passivated surfaces of PPC HPGe detectors \cite{Alpha}. These events deposit energy in the passivated surface, and the generated charge carriers are released to the bulk slowly, which modifies the slope of the falling edge. The DCR parameter is related to the falling edge slope which allows the removal of such background with a signal efficiency of 99\%.

\section{RESULTS}

\hspace{0.65cm}The final background spectrum is generated from 26 kg-yr of data collected between June 2015 and April 2018. After taking into account the cut efficiencies ($\epsilon_{cut}$), the resolution efficiency ($\epsilon_{res}$) and the efficiency for containing the full event energy within the active volume ($\epsilon_{count}$), the exposure in which 0$\nu\beta\beta$ decay can be potentially observed (NT$\epsilon_{cut}$$\epsilon_{res}$$\epsilon_{count}$) is (1.331 $\pm$ 0.063)$\times$10$^{26}$ atom-yr.

\vspace{0.4cm}
The energy range used for background index determination is from 1950 keV to 2350 keV. Based on simulation studies, the background is expected to be flat in this region, with three known gamma peaks as well as the region of interest (ROI: ($Q_{\beta\beta}$ - 5keV, $Q_{\beta\beta}$ + 5keV)) excluded. The background index is found to be (15.4 $\pm$ 2.0) counts/(FWHM t yr) in the vicinity of the ROI  and it assumes the same value in it. The window used for obtaining the half-life is optimized in each subset. The optimized sizes of these windows are around 4.3 keV, which means that 0.653 background events are expected. The best 0$\nu\beta\beta$ limit is given by

\begin{equation}
T^{0\nu}_{1/2} > \ln\left(2\right)\frac{NT\epsilon_{cut}\epsilon_{res}\epsilon_{count}}{S}
\end{equation}

\vspace{0.4cm}
\hspace{-0.7cm}where S is the upper limit on the number of $0\nu\beta\beta$ events. The quoted limit is derived using an unbinned, extended profile likelihood method. While the half-life is common for all datasets, the peak shape parameters and signal efficiencies are constrained to their data set-specific values as Gaussian nuisance terms. The observed lower limit, based on the measured p-value distribution, is \textbf{$T^{0\nu}_{1/2}$ $>$ 2.7 $\times$ 10$^{25}$ yr} at 90\% C.L.\cite{26MJD}. With the matriz element, $g_{A}$, and their uncertainties \cite{26MJD}, the upper limit on $\left\langle m_{\beta\beta}\right\rangle$ is found to be between 200 meV and 430 meV.

\section{ACKNOWLEDGMENTS}
This material is based upon work supported by the U.S. Department of Energy, Office of Science, Office of Nuclear Physics, the Particle Astrophysics and Nuclear Physics Programs ofthe National Science Foundation, and the Sanford Underground Research Facility. It is also funded by the National Science Foundation under Grant No. 1812356.

\bibliographystyle{unsrtnat}
\bibliography{MEDEX_bib}

\begin{thebibliography}{12}
\providecommand{\natexlab}[1]{#1}
\providecommand{\url}[1]{\texttt{#1}}
\expandafter\ifx\csname urlstyle\endcsname\relax
  \providecommand{\doi}[1]{doi: #1}\else
  \providecommand{\doi}{doi: \begingroup \urlstyle{rm}\Url}\fi

\bibitem[et~al. (MAJORANA~Collaboration)(2014)]{MJD}
N.~Abgrall et~al. (MAJORANA~Collaboration).
\newblock \emph{Adv. High Energy Phys.}, \penalty0 (365432), 2014.

\bibitem[et~al.(2009)]{GRETINA}
J.~Anderson et~al.
\newblock \emph{IEEE Trans.Nucl.SSci.}, 56\penalty0 (258), 2009.

\bibitem[et~al. (MAJORANA~Collaboration)(2017)]{Calibration}
N.~Abgrall et~al. (MAJORANA~Collaboration).
\newblock \emph{Nucl.Inst.and Meth.A}, 872:\penalty0 16--22, 2017.

\bibitem[Jordanov and G.F.Knoll(1994)]{Filtered}
V.T. Jordanov and G.F.Knoll.
\newblock \emph{Nucl.Inst.and Meth.A}, 345\penalty0 (337), 1994.

\bibitem[Leo(1994)]{PZ}
W.R. Leo.
\newblock Springer Berlin Heidelberg, 1994.

\bibitem[et~al.(2012)]{CT}
R.~D.~Martin et~al.
\newblock \emph{Nucl.Inst.and Meth.A}, 678:\penalty0 98--104, 2012.

\bibitem[Fano(1947)]{Fano}
U~Fano.
\newblock \emph{Phys.Rev}, 72\penalty0 (26), 1947.

\bibitem[et~al. (MAJORANA~collaboration)(2017)]{DataQ}
J.~Myslik et~al. (MAJORANA~collaboration).
\newblock \emph{Proceedings of TAUP 2017}, arXiv:1711.10550, 2017.

\bibitem[et~al (MAJORANA~collaboration)(2017)]{Muon}
N.Abgrall et~al (MAJORANA~collaboration).
\newblock \emph{Astrpart. Phys.}, 93:\penalty0 70--75, 2017.

\bibitem[et~al. (MAJORANA~Collaboration)(2019{\natexlab{a}})]{MultiSite}
S.~I.~Alvis et~al. (MAJORANA~Collaboration).
\newblock \emph{Phys.Rev.C}, 99\penalty0 (065501), 2019{\natexlab{a}}.

\bibitem[on~behalf~of MAJORANA~Collaboration(2017)]{Alpha}
J.~Gruszko on~behalf~of MAJORANA~Collaboration.
\newblock \emph{J.Phys.Conf.Ser.}, 888, 2017.

\bibitem[et~al. (MAJORANA~Collaboration)(2019{\natexlab{b}})]{26MJD}
S.~I.~Alvis et~al. (MAJORANA~Collaboration).
\newblock \emph{Accepted by PRC}, arXiv:1902.02299, 2019{\natexlab{b}}.

\end{thebibliography}

\end{document}